# Inverse square root level-crossing quantum two-state model

T.A. Ishkhanyan[1,2,3], A.V. Papoyan[2], A.M. Ishkhanyan[1,2], and C. Leroy[3]

[1]Russian-Armenian University, Yerevan, 0051 Armenia
[2]Institute for Physical Research, Ashtarak, 0203 Armenia
[3]Laboratoire Interdisciplinaire Carnot de Bourgogne, UMR CNRS 6303-Université de Bourgogne Franche-Comté, Dijon, 21078 France

We introduce a new unconditionally solvable level-crossing two-state model given by a constant-amplitude optical field configuration for which the detuning is an inverse-square-root function of time. This is a member of one of the five families of bi-confluent Heun models. We prove that this is the only non-classical exactly solvable field configuration among the bi-confluent Heun classes, solvable in terms of finite sums of the Hermite functions. The general solution of the two-state problem for this model is written in terms of four Hermite functions of a shifted and scaled argument (each of the two fundamental solutions presents an irreducible combination of two Hermite functions). We present the general solution, rewrite it in terms of more familiar physical quantities and analyze the time dynamics of a quantum system subject to excitation by a laser field of this configuration.

## 1. Introduction

The two-state problem describes the dynamics of the simplest quantum system with just two energy levels. It is a basic quantum model qualitatively describing diverse important phenomena [1]. However, it presents a rather complicated theoretical problem. Even in its simplest time-dependent formulation, the problem is far from being analytically solved in the general case. As regards several particular exact solutions [2-11], despite their wide applicability, well appreciated during the whole course of the development of quantum mechanics, these solutions were derived, in a sense, in a random way. A systematic treatment of the analytic solutions has started in 1980s [12-15]. However, during quite a long time, the efforts did not provide much progress. As it was shown afterward, the reason was that the applied tools were based on the hypergeometric functions the potential of which is rather limited because of certain intrinsic properties [16-18]. It turned out that already that time the ability of these functions to describe the two-state dynamics was almost exhausted.

Recently, we have reported in total 61 classes of laser field configurations [19-21] for which the two-state problem can be solved in terms of considerably more advanced special functions – the five functions of the Heun class [22-24]. Being different generalizations of hypergeometric functions, the Heun functions suggest several immediate extensions of previously known classes of exactly solvable two-state models. Furthermore, importantly, these functions provide many new classes which do not possess hypergeometric subclasses.

Since the Heun functions involve more parameters, as compared with hypergeometric functions, it is understood that for the new classes of models the amplitude and detuning variation functions of the exciting field will generally be more flexible for different variations (e.g., pulse shape, resonance crossing rate, number of crossings, etc.). The recent research on this subject supports this expectation [19-21,25-27].

Among the two-state models, the most important (and perhaps the most intriguing) ones are the field configurations providing crossings of the resonance (when the external field becomes, at a time point or at several time points, exactly resonant for the two involved quantum states). This is because the resonance crossings may cause a dramatic transfer of populations between the states, not achievable by non-crossing schemes. For this reason, these models may serve as effective tools for quantum control in a prescribed way.

During the last years, applying the Heun functions, we have introduced several new level-crossing models. Though the solution for these models is eventually written in terms of familiar hypergeometric functions, the structure of the solution substantially differs from that for classical hypergeometric models. Namely, the solution is not written through a one-term ansatz involving just one hypergeometric function. Rather, the solution presents an irreducible linear combination of *two* hypergeometric functions. The constant-amplitude model with periodically repeated crossings of the resonance (a member of the general Heun class of models [25]) and the asymmetric Lambert-W model (this belongs to the single-confluent Heun class of models [26]) are representative examples.

In the present paper, we introduce one more level-crossing two-state model the solution for which is of this type. This time, this is a model which belongs to one of the five bi-confluent Heun two-state classes (that one with $k=1$, see [19]). This is a constant-amplitude model for which the detuning of the laser frequency from the transition frequency is an inverse square root function of time. The interaction starts at a finite point of time with an infinite detuning, passes through resonance and asymptotically tends to a finite constant detuning. The general solution for this model is written as a combination, with arbitrary coefficients, of two fundamental solutions each of which presents an irreducible combination of two Hermite functions of shifted and scaled argument. Thus, the general solution of the problem is written in terms of *four* Hermite functions. Examining the initial-value problem, we simplify the general solution and rewrite that in terms of physically appropriate quasi-energies. Finally, we analyze the dynamics of the system subject to excitation with the given model. Both crossing and non-crossing cases are considered.

The level-crossing is a key mechanism of non-adiabatic transitions that play a crucial role in many important phenomena in physics, chemistry, biology [28]. There are basically two formulations of crossing models – time dependent and coordinate dependent. For the time dependent models, which are common in laser excitation processes [1], the essential point is the variation, in time, of the detuning of an exciting laser field. The amplitude of the field may also vary, however, independently of that, the field configuration becomes level-crossing if the detuning of the laser frequency from a transition frequency adopts zero at a time point. The coordinate dependent models, which are common, e.g., in molecular collision physics [29], describe transitions between states whose energies cross as a function of some spatial coordinate. We note that the coordinate dependent problems can be converted to time-dependent ones by supposing a common trajectory and constant velocity [28,29].

The development of the semiclassical theory of non-adiabatic transitions started from the Landau-Zener level-crossing model, which assumes an effective detuning linearly varying in time [2-4]. For this model, the interaction is considered on the full time axis and the detuning infinitely diverges on both asymptotes $t \to \pm\infty$. It has been noticed, however, that in many physical situations the actual conditions essentially differ from these assumptions. For instance, in atomic collision problems, the separation of energy terms is often finite at least at one of the asymptotes. Furthermore, often the interaction should be considered on a semi-axis because of a singularity at a finite time or coordinate point. In these cases, the transition dynamics essentially deviates from the Landau-Zener picture. Some models discussing the effects of these peculiarities have been proposed by Nikitin [7], Demkov and Kunike [9], and Child [10].

The model we introduce has common features with the covalent-ionic interaction model by Child [10] with an effective detuning $\sim C+1/r$ ($r=vt$) describing the inelastic scattering of alkali atoms from neutral targets. Like this model, our model involves a finite detuning on an asymptote and has a singularity at the origin. Since the Child model belongs to the same confluent hypergeometric class [16], it has much in common with the Landau-Zener dynamics [10]. However, it also reveals important differences such as the non-zero limit of the non-transition probability at infinitely increasing coupling strengths [30]. Our model reveals more deviation from the Landau-Zener model, e.g., it shows more expressed saturation behavior. We expect that this peculiarity can be observed, e.g., in Dirac materials where a field involving inverse-square-root dependence is created at the edge of a metallic contact with graphene due to a gate voltage [31].

## 2. The field configuration

The semiclassical time-dependent two-state problem is written as a system of coupled first-order differential equations for probability amplitudes $a_1(t)$ and $a_2(t)$ of given two states of a quantum system. If the system is driven by an external field with amplitude modulation $U(t) > 0$ and phase modulation $\delta(t)$, this system is written as [1]

$$i\frac{da_1(t)}{dt} = U(t)e^{-i\delta(t)}a_2(t), \qquad (1)$$

$$i\frac{da_2(t)}{dt} = U(t)e^{+i\delta(t)}a_1(t) \qquad (2)$$

Note that this form assumes a lossless interaction, that is, we do not consider the dissipation. This system is equivalent to the following second-order linear ordinary differential equation:

$$\frac{d^2 a_2}{dt^2} + \left(-i\delta_t(t) - \frac{U_t(t)}{U(t)}\right)\frac{da_2}{dt} + U^2(t) a_2 = 0, \qquad (3)$$

where $\delta_t$ and $U_t$ refer to the time derivatives of corresponding functions.

The external laser field-configuration model we introduce is the following constant-amplitude level-crossing two-state model with the detuning variation function given as an *inverse square root* function of time:

$$U(t) = U_0, \quad \delta_t(t) = \Delta_0 + \frac{\Delta_1}{\sqrt{t}}. \qquad (4)$$

These functions are shown in figure 1. The model is level-crossing if $\Delta_0$ and $\Delta_1$ are of opposite signs: $\Delta_0 \Delta_1 < 0$.

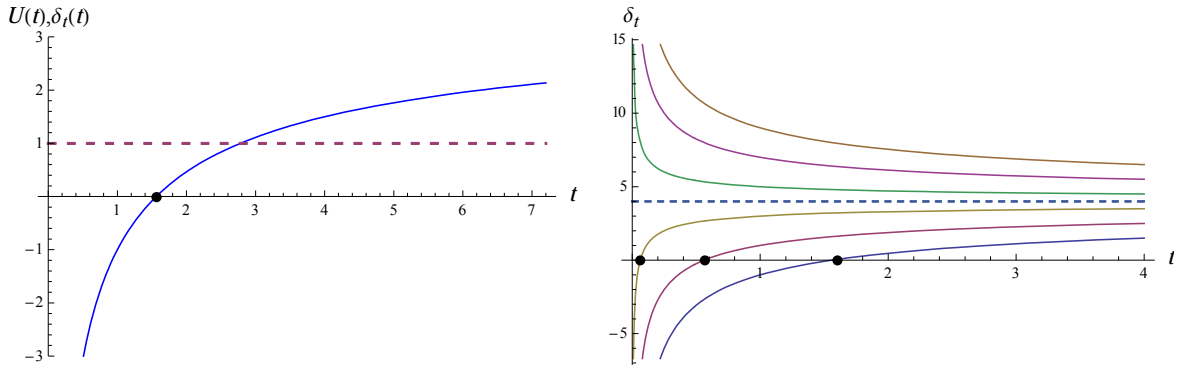

Fig. 1. Two-state level-crossing model (4); filled circles indicate the resonance crossing time points. Left figure: $U_0 = 1, \Delta_0 = 4, \Delta_1 = -5$. The dashed line is the Rabi frequency $U(t) = U_0$ and the solid line presents the detuning $\delta_t(t)$. Right figure: $\delta_t(t)$ for $\Delta_0 = 4$ and $\Delta_1 = 5, 3, 1, -1, -3, -5$ (from top to bottom). The dashed line presents $\delta_t$ for $\Delta_1 = 0$ (asymptotic limit of $\delta_t$ at $t \to \infty$ for any $\Delta_1$).

The resonance crossing happens at time-point

$$t_0 = \frac{\Delta_1^2}{\Delta_0^2}, \quad \Delta_0 \Delta_1 < 0. \tag{5}$$

In the vicinity of this point, the detuning behaves as an approximately linear function of time. The Taylor series expansion reads

$$\delta_t(t) = \frac{\Delta_0^3}{2\Delta_1^2}(t-t_0) + ..., \tag{6}$$

hence, the effective Landau-Zener parameter, which controls the interaction regime at the crossing (see [1-3]), is given as

$$\Lambda = \frac{U_0^2 \Delta_1^2}{4\Delta_0^3}. \tag{7}$$

The fast sweep through the resonance corresponds to $\Lambda \ll 1$, and the sweep is slow if $\Lambda \gg 1$.

## 3. The solution of the two-state problem

It is readily checked by direct substitution that equation (3) for this field configuration admits a *fundamental* solution, involving two non-integer order Hermite functions of a scaled and shifted argument [32], written as

$$a_{2F}(t) = e^{\alpha_0 \sqrt{t} + \alpha_2 t/2} \left( A H_{-\alpha/\varepsilon}(y) + H_{-\alpha/\varepsilon - 1}(y) \right) \tag{8}$$

where

$$A = S\sqrt{\frac{-\varepsilon}{2}} \frac{\delta - q}{\alpha}, \quad y = S\sqrt{\frac{-\varepsilon}{2}} \left( \sqrt{t} + \frac{\delta}{\varepsilon} \right), \tag{9}$$

$$(\delta, \varepsilon, \alpha, q) = \left( 2(\alpha_0 - i\Delta_1), 2(\alpha_2 - i\Delta_0), \alpha_0(\alpha_0 - 2i\Delta_1), \alpha_0 \right), \tag{10}$$

$$\alpha_0 = i\Delta_1 \left( 1 \pm \frac{\Delta_0}{\sqrt{4U_0^2 + \Delta_0^2}} \right), \quad \alpha_2 = i\left( \Delta_0 \pm \sqrt{4U_0^2 + \Delta_0^2} \right). \tag{11}$$

The derivation of the model and this solution are presented in the **Appendix**.

Here the auxiliary parameter $S$ may adopt values $S = +1$ or $S = -1$. We note that each of these values produces a linearly independent fundamental solution. Besides, any sign, plus or minus, is applicable for $\alpha_{0,2}$. For definiteness, we choose the *minus* sign for both alphas. The general solution of the problem is then written as

$$a_2(t) = C_1 a_{2F}|_{S=-1} + C_2 a_{2F}|_{S=+1}, \tag{12}$$

where $C_{1,2}$ are arbitrary constants.

Taking into account the basic identity for the Hermite functions [32]

$$\frac{dH_\nu(y)}{dy} = 2\nu H_{\nu-1}(y),\qquad(13)$$

the fundamental solution (8) can be rewritten in the following equivalent form:

$$a_{2F} = e^{i\lambda\left(t+\frac{\Delta_1\sqrt{t}}{\lambda-\Delta_0/2}\right)}\left(F + \frac{i\sqrt{t}(\Delta_0-2\lambda)}{\Delta_1(\Delta_0-\lambda)}\frac{dF}{dt}\right),\qquad(14)$$

with

$$F(t) = H_{-\frac{2i\Delta_1^2 U_0^2}{(\Delta_0-2\lambda)^3}}\left(\sqrt{i(\Delta_0-2\lambda)}\left(\sqrt{t}+\frac{\Delta_1\Delta_0}{(\Delta_0-2\lambda)^2}\right)\right),\qquad(15)$$

where the parameter $\lambda$ is one of the two *quasi-energies* [33] for $t\to+\infty$:

$$\lambda_{1,2} = \frac{\Delta_0}{2}\pm R,\quad R = \sqrt{\frac{\Delta_0^2}{4}+U_0^2}\qquad(16)$$

(note that $\lambda_{1,2}$ do not depend on $\Delta_1$). With this, the general solution of equation (3) is rewritten in terms of quasi-energies at the end of the interaction process:

$$a_2(t) = C_1 a_{2F}\big|_{\lambda\to\lambda_1} + C_2 a_{2F}\big|_{\lambda\to\lambda_2}.\qquad(17)$$

## 4. The scattering problem for $t\to+\infty$

The structure of solution (17) is such that it is straightforward to study the *scattering* problem posed for $t\to+\infty$. To formulate this problem in the appropriate way, one should first examine the behavior of the quasi-energy states, respectively, with $\lambda_1$ and $\lambda_2$, at adiabatically slowly switching off the interaction ($U_0\to 0$). The Rabi problem [34] for a laser field with $U=U_0=\text{const}$ and $\delta_t=\Delta_0=\text{const}$ reads

$$i\frac{da_1}{dt} = U_0 e^{-i\Delta_0 t}a_2,\quad i\frac{da_2}{dt} = U_0 e^{+i\Delta_0 t}a_1(t),\qquad(18)$$

or (see (3))

$$\frac{d^2 a_2}{dt^2} - i\Delta_0\frac{da_2}{dt} + U_0^2 a_2 = 0.\qquad(19)$$

The solution of this equation is given as

$$a_2 = C_1 e^{i\lambda_1 t} + C_2 e^{i\lambda_2 t}\qquad(20)$$

where $C_{1,2}$ are arbitrary constants and $\lambda_{1,2}$ are given by equation (16). For the first level's probability amplitude $a_1$, from the second equation (18), we have

$$a_1 = -C_1\frac{\lambda_1}{U_0}e^{i(-\Delta_0+\lambda_1)t} - C_2\frac{\lambda_2}{U_0}e^{i(-\Delta_0+\lambda_2)t}.\qquad(21)$$

The behavior of this solution at switching off the interaction depends on the sign of $\Delta_0$. For definiteness, we suppose $\Delta_0 > 0$. Then,

$$\lambda_1 = \frac{\Delta_0}{2} + R \bigg|_{U_0 \to 0} \to \Delta_0, \tag{22}$$

$$\lambda_2 = \frac{\Delta_0}{2} - R \bigg|_{U_0 \to 0} \to 0. \tag{23}$$

Now let, for a moment, $C_2 = 0$. The normalization condition gives $C_1 = -U_0 / \sqrt{U_0^2 + \lambda_1^2}$. Hence, if the Rabi frequency $U_0$ adiabatically slowly goes to zero, we have $(a_1, a_2) \to (1, 0)$. This means that the quasi-energy $\lambda_1 = \Delta_0 / 2 + R$ is the one which stands for the quasi-energy state originating from the *first* energy level at adiabatically slow inclusion of the interaction. Similarly, the analysis of the auxiliary case $C_1 = 0$ shows that the quasi-energy state with $\lambda_2 = \Delta_0 / 2 - R$ originates from the second energy level. We stress that these speculations apply to the case of positive $\Delta_0 > 0$ (for negative $\Delta_0 < 0$, $\lambda_1$ and $\lambda_2$ change roles).

Now, returning to the excitation with the inverse-square-root field configuration (4), consider the asymptotes at the infinity for the two fundamental solutions (14) involved in the general solution (17). Using the properties of the Hermite function [32], it can be shown that these asymptotes are

$$a_{2F}^{(\lambda_1)}\bigg|_{t \to +\infty} \approx e^{\frac{\pi \Delta_1^2 U_0^2}{16 R^3}} e^{i\left(\lambda_1 t + \frac{\Delta_1 \lambda_1}{R}\sqrt{t} + \frac{\Delta_1^2 U_0^2}{8R^3}\ln(8Rt)\right)}, \tag{24}$$

$$a_{2F}^{(\lambda_2)}\bigg|_{t \to +\infty} \approx e^{\frac{4\pi \Delta_1^2 U_0^2}{16 R^3}} e^{i\left(\lambda_2 t - \frac{\Delta_1 \lambda_2}{R}\sqrt{t} - \frac{\Delta_1^2 U_0^2}{8R^3}\ln(2Rt)\right)}, \tag{25}$$

so that

$$a_2\big|_{t \to +\infty} \sim C_1 e^{i\lambda_1 t} + C_2 e^{i\lambda_2 t}. \tag{26}$$

Consider now the scattering problem when the system at $t \to +\infty$ ends in the *first* quasi-energy state with the quasi-energy $\lambda_1$. We note that in general the field configuration (4) is not adiabatically slow, hence, considerable differences are expected as compared with the Rabi model. The first quasi-energy state is achieved if $C_2 = 0$. Then,

$$a_2\big|_{t \to +\infty} \approx C_1 e^{\frac{\pi \Delta_1^2 U_0^2}{16 R^3}} e^{i\lambda_1 t}. \tag{27}$$

The corresponding asymptote of $a_1$ reads

$$a_1\big|_{t\to+\infty} = \frac{ia_2'}{Ue^{i(\Delta_0 t+2\Delta_1\sqrt{t})}}\bigg|_{t\to+\infty} \approx -C_1\frac{\lambda_1}{U_0}e^{\frac{\pi\Delta_1^2 U_0^2}{16R^3}}e^{-i(\lambda_2 t+2\Delta_1\sqrt{t})}. \tag{28}$$

The normalization condition $|a_1|^2 + |a_2|^2 = 1$ now gives the value of the coefficient $C_1$:

$$C_1 = \frac{U_0}{\sqrt{U_0^2 + \lambda_1^2}} e^{-\frac{\pi\Delta_1^2 U_0^2}{16R^3}}. \tag{29}$$

The probability amplitude of the second level at the beginning of interaction at $t=0$ is

$$a_2(0) = C_1\left(H_\nu(\xi) + \left(-\xi + \sqrt{\xi^2 - 2\nu}\right)H_{\nu-1}(\xi)\right), \tag{30}$$

where
$$(\nu,\xi) = \left(\frac{iU_0^2\Delta_1^2}{4R^3}, \frac{(1-i)\Delta_0\Delta_1}{4R^{3/2}}\right). \tag{31}$$

It can be shown that this function never adopts exact zero (see below). This means that, starting from the second energy level at $t=0$, it is impossible to achieve a final state with a purely first quasi-energy state, that is, without exciting the second quasi-energy component.

The 3D plot of the occupation probability of the first energy level $p_1(0) = 1 - |a_2(0)|^2$ is shown in figure 2. This figure well illustrates the role of the level crossing. If $\Delta_1 > 0$, we have a non-crossing excitation. In this case, the initial population of the first level is almost one (upper part of figure 2); the system starts from the first energy level and ends in the first quasi-energy state. In contrary, if the detuning crosses the resonance, in order to end in the first quasi-energy state, the system should start from the second level (lower left corner) or a superposition state (the region around the diagonal $\Delta_1 = -\Delta_0$).

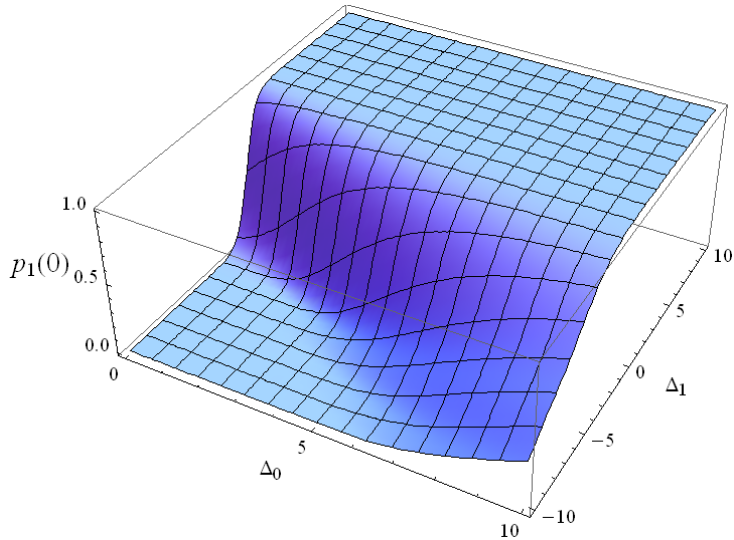

Fig.2. Occupation probability of the first level at the beginning of interaction, $U_0 = 1$.

## 5. Probability amplitude at $t=0$.

Examining the probability amplitude (30) for $t=0$:

$$a_2(0) = C_1\left(H_\nu(\xi) + \left(-\xi + \sqrt{\xi^2 - 2\nu}\right)H_{\nu-1}(\xi)\right), \qquad (32)$$

we observe that the interaction process is controlled by two dimensionless real parameters:

$$(\nu_0, \xi_0) = \left(\frac{U_0^2 \Delta_1^2}{4R^3}, \frac{\Delta_0 \Delta_1}{4R^{3/2}}\right). \qquad (33)$$

With this, we have

$$\nu = i\nu_0, \quad \xi = (1-i)\xi_0, \qquad (34)$$

and the constant $C_1$ is rewritten as

$$C_1 = \frac{e^{-\frac{\pi\nu_0}{4}}}{\sqrt{2}}\sqrt{1 + \frac{\xi_0}{\sqrt{\nu_0 + \xi_0^2}}}. \qquad (35)$$

The behavior of $\nu_0$ and $\xi_0$ as functions of $U_0$ is shown in figure 3. We observe that the parameter $\nu_0$ is always positive, and, for a level-crossing configuration with $\Delta_0 > 0$ and $\Delta_1 < 0$, $\xi_0$ is negative. The asymptotes of the two parameters for $U_0 \to 0$ and $U_0 \to \infty$ read

$$\nu_0\big|_{U_0 \to 0} \approx \frac{2\Delta_1^2}{\Delta_0^3}U_0^2, \quad \nu_0\big|_{U_0 \to \infty} \approx \frac{\Delta_1^2/4}{U_0}, \qquad (36)$$

$$\xi_0\big|_{U_0 \to 0} \approx \frac{\Delta_1}{\sqrt{2\Delta_0}}, \quad \xi_0\big|_{U_0 \to \infty} \approx \frac{\Delta_0 \Delta_1/4}{U_0^{3/2}}. \qquad (37)$$

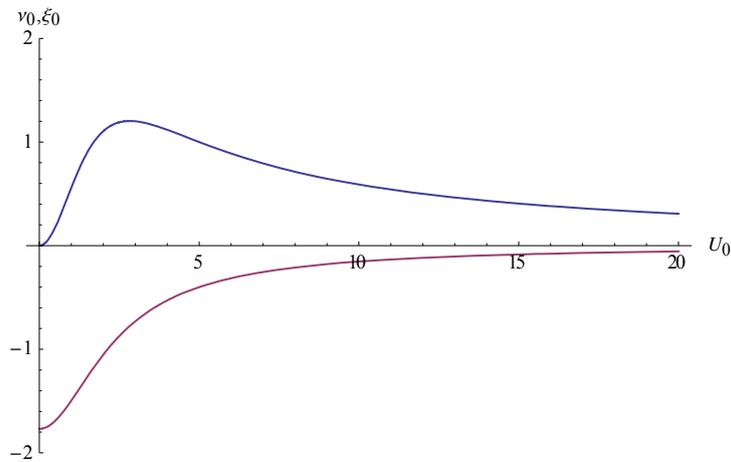

Fig. 3. The parameters $\nu_0$ (upper curve) and $\xi_0$ (lower curve) as functions of $U_0$. $\Delta_0 = 4$, $\Delta_1 = -5$.

To examine the weak field limit $U_0 \to 0$, we note that in this limit $\nu_0$ becomes proportional to the generalized Landau-Zener parameter $\Lambda$ (see equation (7)), while $\xi_0$ tends to a constant. Then, noting that $\nu = -2\xi^2 U_0^2 / \Delta_0^2$, one may expand the Hermite functions involved in equation (32) for small $U_0$ to derive the approximation

$$a_2(0) \approx C_1 \left( 1 + \frac{\sqrt{\pi}}{2} \left( \sqrt{\xi^2} - \xi \right) e^{\xi^2} \mathrm{erfc}(\xi) \right), \qquad (38)$$

where $\mathrm{erfc}(\xi)$ is the complementary error function: $\mathrm{erfc}(\xi) = 1 - \mathrm{erf}(\xi)$ [24].

For the opposite limit of strong field, $U_0 \to \infty$, we note that both $\nu_0$ and $\xi_0$ tend to zero. With this, the standard power-series expansion of the Hermite functions [32] produces the zero-order approximation

$$a_2(0) \approx C_1 \sqrt{\pi}\, 2^{\nu} \left( \frac{\sqrt{-\nu}}{\sqrt{2}\, \Gamma\!\left(1 - \dfrac{\nu}{2}\right)} + \frac{1}{\Gamma\!\left(\dfrac{1-\nu}{2}\right)} \right). \qquad (39)$$

where $\Gamma$ is the Euler gamma function.

The presented approximations, applied for calculation of the occupation probability $p_2(0) = |a_2(0)|^2$, are shown in figure 4. A short examination of the approximations reveals that the probability $p_2(0)$ does not adopt zero unless $U_0 = 0$.

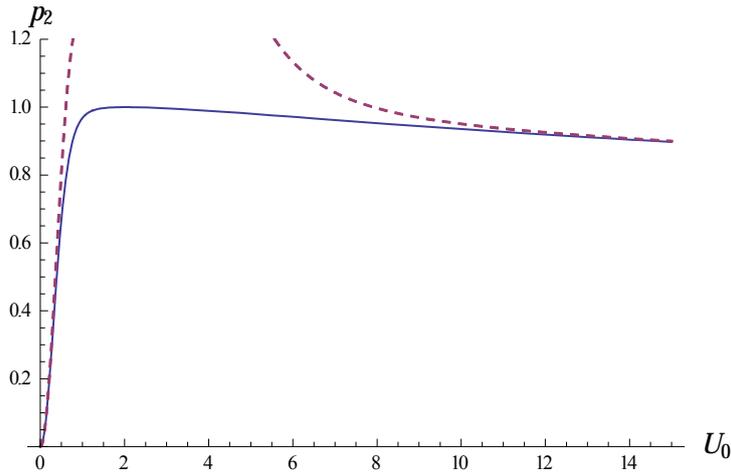

Fig. 4. Approximations (38) for $U_0 \to 0$ and (39) for $U_0 \to \infty$. The solid line presents the exact result for $p_2(0) = |a_2(0)|^2$. $\Delta_0 = 4$, $\Delta_1 = -5$.

## 6. Discussion

Thus, we have presented a new constant-amplitude level-crossing two-state model with the detuning varying in time as the inverse square root function. This is an exactly integrable model derived by the termination of the Hermite-function expansion of the bi-confluent Heun function on the second term. The general solution of the problem is written in terms of two independent fundamental solutions each of which presents an irreducible linear combination of two Hermite functions of a shifted and scaled argument. Rewriting the solution in terms of quasi-energies, we have discussed the scattering problem when at the end of the interaction the system occupies only the first quasi-energy state. Discussing both crossing and non-crossing scenarios, we have seen that the model confirms the tremendous difference between the two cases.

The model we have described can be relevant in different physical situations. For instance, the calculations using density function theory in the Thomas-Fermi approximation show that the contact-induced electrostatic potential caused by a gate voltage at the edge of a graphene strip is of the functional form involving the inverse-square-root: $V \sim V_0 + V_1/\sqrt{x}$ [31,38]. It is then expected that a level-crossing excitation of the type we have modeled can occur for two-level quantum systems slowly moving ($x \sim vt$) in such a field.

**Acknowledgments**

The research was supported by the Russian-Armenian (Slavonic) University at the expense of the Ministry of Education and Science of the Russian Federation, the Armenian Science Committee (SC Grant No. 18T-1C276), and the Armenian National Science and Education Fund (ANSEF Grant No. PS-5701). T.A. Ishkhanyan acknowledges the support from the French Embassy in Armenia for a doctoral grant and thanks the Foundation for Armenian Science and Technology for a PhD Fellowship grant.

## Appendix. Derivation of the inverse-square-root model

For completeness of the treatment, we present the comprehensive derivation of the inverse-square-root model and the solution (8)-(11) of the two-state problem for this model.

It has been previously shown in [21] that equation (3) is reduced to the bi-confluent Heun equation [22]

$$\frac{d^2u}{dz^2} + \left(\frac{\gamma}{z} + \delta + \varepsilon z\right)\frac{du}{dz} + \frac{\alpha z - q}{z}u = 0 \qquad (40)$$

for five infinite classes of optical field configurations given as

$$U(t) = U_0^* z^k \frac{dz}{dt}, \tag{41}$$

$$\delta_t(t) = \left(\frac{\delta_1}{z} + \delta_0 + \delta_2 z\right)\frac{dz}{dt}. \tag{42}$$

Here $k = -1, -1/2, 0, 1/2, 1$, $z(t)$ is an arbitrary complex-valued function of time, and the parameters $U_0^*$ and $\delta_{0,1,2}$ are complex constants that should be chosen so that $U(t)$ and $\delta_t(t)$ are real. Since the parameters are arbitrary, the classes are four-parametric in general.

The solution of the two-state problem is explicitly written as [21]

$$a_2 = z^{\alpha_1} e^{\alpha_0 z + \frac{\alpha_2}{2} z^2} H_B(\gamma, \delta, \varepsilon; \alpha, q; z), \tag{43}$$

where $H_B$ is the bi-confluent Heun function the parameters of which are given as

$$\gamma = 2\alpha_1 - i\delta_1 - k, \quad \delta = 2\alpha_0 - i\delta_0, \quad \varepsilon = 2\alpha_2 - i\delta_2, \tag{44}$$

$$\alpha = \alpha_0(\alpha_0 - i\delta_0) + \alpha_1(2\alpha_2 - i\delta_2) + \alpha_2(1 - k - i\delta_1) + Q''(0)/2, \tag{45}$$

$$q = \alpha_0(k + i\delta_1) - \alpha_1(2\alpha_0 - i\delta_0) - Q'(0) \tag{46}$$

with

$$Q(z) = U_0^{*2} z^{2k+2} \tag{47}$$

and

$$\alpha_0 \varepsilon - i\alpha_2 \delta_0 + Q'''(0)/3! = 0, \tag{48}$$

$$\alpha_1^2 - \alpha_1(1 + k + i\delta_1) + Q(0) = 0, \tag{49}$$

$$\alpha_2^2 - i\alpha_2 \delta_2 + Q^{(4)}(0)/4! = 0. \tag{50}$$

Furthermore, it has been shown in [35] that for a non-zero $\varepsilon$ the bi-confluent Heun function involved in solution (43) can be expanded in terms of the Hermite functions of a shifted and scaled argument:

$$H_B = \sum_{n=0}^{\infty} c_n H_{n+\gamma-\alpha/\varepsilon}\left(\pm\sqrt{-\varepsilon/2}\,(z + \delta/\varepsilon)\right), \tag{51}$$

where the coefficients $c_n$ obey the three-term recurrence relation

$$R_n c_n + Q_{n-1} c_{n-1} + P_{n-2} c_{n-2} = 0 \tag{52}$$

with

$$R_n = \frac{\sqrt{2}}{\sqrt{-\varepsilon}} n\left(-\alpha + (\gamma + n)\varepsilon\right), \tag{53}$$

$$Q_n = \mp\left(q + (\gamma + n)\delta\right), \tag{54}$$

$$P_n = \frac{(\gamma + n)\varepsilon}{\sqrt{-2\varepsilon}}. \tag{55}$$

Here the signs $\mp$ in equation (54) for $Q_n$ refer to the choices $s_0 = \pm\sqrt{-\varepsilon/2}$, respectively.

These series may terminate producing closed-form finite-sum solutions. This happens if two successive coefficients vanish for some $n = N = 0, 1, 2, ...$, i.e., if $c_{N+1} = c_{N+2} = 0$ while $c_N \neq 0$. It follows from equation $c_{N+2} = 0$ that the termination is possible only if $P_N = 0$. Since $\varepsilon$ is not zero, this condition is satisfied if $\gamma = -N$. The remaining equation $c_{N+1} = 0$ then presents a polynomial equation of the degree $N+1$ for the accessory parameter $q$. This gives $N+1$ values of $q$ for which the termination of the series occurs. Here are the necessary equations for $N = 0$ and 1:

$$\gamma = 0: \quad q = 0. \tag{56}$$

$$\gamma = -1: \quad q^2 - \delta q + \alpha = 0. \tag{57}$$

Thus, the termination of series (51) imposes two restrictions on the parameters of the involved bi-confluent Heun function. One of these restrictions is imposed on a characteristic exponent of the finite singularity of the bi-confluent Heun equation, while the second one presents a polynomial equation for the accessory parameter. The condition imposed on a characteristic exponent generally leads to *conditionally* integrable models for which the involved field-parameters do not vary independently. However, there are four remarkable exceptions which result in exactly solvable models. The first three of these cases reproduce the known models of Landau-Zener [2-5], Nikitin [7] and Demkov [8], while the fourth one presents a new result. In this case, the detuning function behaves as the inverse square root function at the origin and goes to a constant detuning at infinity.

To see the details, consider the termination condition given by equation

$$\gamma = -N, \tag{58}$$

which is a restriction imposed on the characteristic exponent of the singularity of the bi-confluent Heun equation located at the origin $z = 0$. With this condition, we examine the equations for the exponent $\alpha_1$ of the pre-factor of the solution (43), that is, the first equation (44) and equation (49):

$$\gamma = 2\alpha_1 - i\delta_1 - k, \tag{59}$$

$$\alpha_1^2 - \alpha_1(1 + k + i\delta_1) + Q(0) = 0, \tag{60}$$

Eliminating $\alpha_1$, we have the equation

$$(\gamma + k + i\delta_1)(\gamma - k - 2 - i\delta_1) = -4Q(0). \tag{61}$$

The function $Q(z) = U_0^{*2} z^{2k+2}$ depends only on the amplitude parameter $U_0^*$, hence, $Q(0)$ is a function of $U_0^*$ only. It is then understood that if $\gamma$ is a fixed constant, we have an equation for which the left-hand side depends on the detuning parameter $\delta_1$, while the right-hand side depends on the amplitude parameter $U_0^*$. The conclusion is that in general the parameters $\delta_1$ and $U_0^*$ are not independent, they are related by equation (61). Hence, the models solvable in terms of linear combinations of the Hermite functions of shifted and scaled argument are in general *conditionally* integrable.

Exceptional are the cases for which both sides of equation (61) identically vanish. For these cases, from the condition $Q(0) = 0$, should be $k \neq -1$. Besides, it follows from the left-hand side of equation (61) that the detuning parameter $\delta_1$ is either a fixed number or should vanish. In the first case one obtains *dissipative* conditionally integrable models (see, e.g., [36]), while the second case, for which $\delta_1 = 0$, leads to *exactly* solvable ones.

For the latter models, equation (61) reads

$$(\gamma + k)(\gamma - k - 2) = 0. \tag{62}$$

Since $\gamma$ is a non-positive integer and $k$ may adopt only four integer or half-integer values: $k = -1/2, 0, 1/2, 1$ ($k \neq -1$), we conclude that exactly solvable models are achieved only if $\gamma = k = 0$ or $\gamma = -1 \cup k = 1$. The first case $\gamma = 0$ is accompanied with the equation $q = 0$ (see equation (56)). This case is not of much interest because then the bi-confluent Heun equation directly reduces to the confluent hypergeometric equation and, as a result, one obtains the three known confluent hypergeometric models by Landau-Zener, Nikitin and Crothers-Hughes (see the details in [37]).

A new result is achieved if one considers the case when the series expansion is terminated on the second term. In this case, we have $\gamma = -1$ and the second condition for termination is given by equation (57):

$$q^2 - \delta q + \alpha = 0. \tag{63}$$

We recall that for exact solvability it additionally should be $k = 1$ and $\delta_1 = 0$. From the first equation (44), we have $\alpha_1 = 0$. With this, it is readily checked that equation (63) is satisfied for all other parameters being arbitrary. Thus, we derive a new *three-parametric* class of exactly solvable bi-confluent-Heun two-state models:

$$U(t) = U_0^* \cdot z \frac{dz}{dt}, \tag{64}$$

$$\delta_t(t) = (\delta_0 + \delta_2 z) \frac{dz}{dt}. \tag{65}$$

By choosing the transformation of the independent variable as

$$z(t) = \sqrt{t}, \tag{66}$$

we arrive at the *constant-amplitude inverse square root* level-crossing model (4):

$$U(t) = U_0, \tag{67}$$

$$\delta_t(t) = \Delta_0 + \frac{\Delta_1}{\sqrt{t}}, \tag{68}$$

where we have put $U_0^* = 2U_0$, $\delta_0 = 2\Delta_1$, $\delta_2 = 2\Delta_0$. With this, the solution (43)-(50) is directly rewritten as (8)-(11).